\documentclass[a4paper]{article}
\usepackage{amscd,amsmath,amssymb}
\usepackage[usenames, dvipsnames]{color}

\newcommand{\R}{\mathbb R}

\newcommand{\p}[1]{(\ref{#1})}

\newcommand{\bQ}{{\overline Q}{}}

\newcommand{\mN}{\mathcal{N}}

\newcommand{\bpsi}{{\bar\psi}{}}

\newcommand{\bu}{{\bar u}}

\newcommand{\sfrac}[2]{{\textstyle\frac{#1}{#2}}}
\newcommand{\und}{\qquad\textrm{and}\qquad}
\renewcommand{\=}{\ =\ }

\newcommand{\be}{\begin{equation}}
\newcommand{\ee}{\end{equation}}
\newcommand{\bea}{\begin{eqnarray}}
\newcommand{\eea}{\end{eqnarray}}

\newcommand{\ba}{\begin{array}} \newcommand{\ea}{\end{array}}

\def\im{{\rm i}}
\def\ep{{\rm e}}

\newcommand{\nn}{\nonumber}

\begin{document}

\pagenumbering{gobble}

\begin{flushright}
\phantom{.}
\end{flushright}

\vspace{2cm}

\begin{center}
{\LARGE\bf SU(2$|$1) supersymmetric mechanics on curved spaces}
\end{center}

\vspace{1cm}

\begin{center}
\mbox{\Large\bf  
Nikolay~Kozyrev${}^{a}$, 
Sergey~Krivonos${}^{a}$, 
Olaf~Lechtenfeld${}^{b}$,
Anton~Sutulin${}^{a}$
}
\end{center}

\vspace{1cm}

\begin{center}
${}^a$ {\it
Bogoliubov  Laboratory of Theoretical Physics, JINR,
141980 Dubna, Russia}

\vspace{0.2cm}

${}^b$ {\it
Institut f\"ur Theoretische Physik and Riemann Center for Geometry and Physics \\
Leibniz Universit\"at Hannover,
Appelstrasse 2, 30167 Hannover, Germany}

\vspace{0.2cm}

{\tt nkozyrev, krivonos, sutulin @theor.jinr.ru, lechtenf@itp.uni-hannover.de}

\end{center}

\vspace{2cm}

\begin{abstract}\noindent
We present  SU$(2|1)$ supersymmetric mechanics on  $n$-dimensional Riemannian manifolds
within the Hamiltonian approach. The structure functions including prepotentials entering the 
supercharges and the Hamiltonian obey extended curved WDVV equations specified by the manifold's 
metric and curvature tensor. We consider the most general $u(2)$-valued prepotential, which contains 
both types (with and without spin variables), previously considered only separately. For the case of 
real K\"{a}hler manifolds we construct all possible interactions. For isotropic ($so(n)$-invariant) spaces 
we provide admissible prepotentials for any solution to the curved WDVV equations. 
All known one-dimensional SU$(2|1)$ supersymmetric models are reproduced.
\end{abstract}

\newpage
\pagenumbering{arabic}
\setcounter{page}{1}

\section{Introduction}
One of the interesting features of $\mN{=}\,4$ supersymmetric mechanics is its relation with the 
Witten--Dijkgraaf--Verlinde--Verlinde (WDVV) equations \cite{WDVV}.  
The most natural appearance of the WDVV equations  is seen at the component level.
As was first demonstrated in \cite{Wyllard}, on the $(2n{+}4)$-dimensional phase space 
$\bigl\{ x^i, p_j, \psi^{ai},\bpsi^j_b\bigr\}$, with $i,j=1,\ldots,n$ and $a,b=1,2$, 
the simplest ansatz for the $\mN{=}\,4$ supercharges $Q^a$ and $\bQ_a$, 
\be\label{introQ}
Q^{a} \= p_i \psi^{ia} + \im F^{(0)}_{ijk} \psi^{ib}\psi_b^j \bpsi^{ka} \und
\bQ_a \= p_i \bpsi^{i}_a + \im F^{(0)}_{ijk} \bpsi^{i}_b\bpsi^{jb} \psi^{k}_a \ ,
\ee
yields the WDVV equations
\be\label{introWDVV}
F^{(0)}_{ijm} \delta^{nm}F^{(0)}_{kln}-F^{(0)}_{ilm} \delta^{nm}F^{(0)}_{kin} \=0
\qquad \textrm{with}\qquad
F^{(0)}_{ijk} =\partial_i \partial_j \partial_k F^{(0)}(x)
\ee
for totally symmetric structure functions $F^{(0)}_{ijk}$, 
if one requires the supercharges to obey the $\mN{=}\,4$ super Poincar\'{e} algebra
\be\label{introN4P}
\left\{ Q^{a}, \bQ_b \right\} = \sfrac{\im}{2} \delta^a_b H\ , \quad 
\left\{ Q^{a}, Q^{b} \right\}=0\ , \quad
\left\{ \bQ_a, \bQ_b \right\}=0\ .
\ee
The evaluation of the brackets in \p{introN4P} assumed the standard Dirac brackets 
between the basic variables,
\be\label{introDB}
\left\{ x^i , p_j\right\}= \delta^i_j \und 
\bigl\{ \psi^{ia}, \bpsi^j_b\bigr\} = \sfrac{\im}{2} \delta^a_b \delta^{ij}\ .
\ee

The simplest form~\p{introQ} of the supercharges does not produce (classically) any potential term 
in the Hamiltonian $H$. To generate physically interesting systems, the supercharges have to be 
extended by terms linear in the fermionic variables.
Such linear terms come with new structure functions, so-called prepotentials, 
which obey differential equations extending the curved WDVV ones.
Prepotentials come in two variants, called $W$ and $U$. The first one is associated with
a $u(1)$ subalgebra of the $u(2)$ R-symmetry algebra, 
the second one with an $su(2)$ subalgebra. The latter
requires the introduction of semi-dynamical spin variables~\cite{FIL}.
Examples of such constructions can be found in \cite{Wyllard, BGL, GLP, GLP1,KL1} 
and references therein. 

So far we discussed $\mN{=}\,4$ supersymmetric mechanics on the Euclidian space~$\R^n$.
Recently~\cite{cWDVV,N4mech}, the structure given by \p{introQ}--\p{introDB} was generalized 
to $\mN{=}\,4$ supersymmetric mechanics on arbitrary Riemann spaces, rendering it covariant
under general coordinate transformations. In this case, the WDVV equations \p{introWDVV} 
are superseded by the `curved WDVV equations'~\cite{cWDVV}
\be\label{introcWDVV}
\nabla_i F_{jkm} = \nabla_j F_{ikm} \und
F_{ikp}g^{pq}F_{jmq}-F_{jkp}g^{pq}F_{imq}+R_{ijkm} \=0
\ee
involving the Riemann tensor~$R$ of the Riemannian manifold.
Simultaneously, the conditions on the prepotentials $W$ or $U$ entering the supercharges 
have been covariantized~\cite{N4mech}. 

Another generalization of $\mN{=}\,4$ supersymmetric mechanics has been proposed by
Smilga~\cite{Smilga}, by adding R-symmetry generators in the right-hand side of the basic 
commutators $\left\{ Q^a, \bQ_b\right\} =\frac{\im}{2} \delta^a_b H$.
This step deforms the $\mN{=}\,4$ super Poincar\'{e} algebra to an $su(2|1)$ algebra~\cite{Smilga}. 
A systematic study of one-dimensional SU$(2|1)$ supersymmetric mechanics has been conducted 
in~\cite{Sidorov1, Sidorov1a,Sidorov1b,Sidorov2} using the superspace approach. 

Our main goal is to construct $n$-dimensional SU$(2|1)$ supersymmetric 
mechanics with a $(2n{+}4n)$-dimensional phase space over an arbitrary Riemannian manifold
within the Hamiltonian approach.\footnote{
Particular cases of $\mN{=}\,2,4$ supersymmetric mechanics with weak supersymmetry 
and $(4n{+}4n)$-dimensional phase spaces have been considered in~\cite{BN}.}
In Section 2 we introduce generalized Poisson brackets which are general coordinate covariant,  
write down the most general ansatz for the supercharges (linear and cubic in the fermionic variables), 
and analyze the conditions on the structure functions. 
These determine the structure functions and the explicit structure of the Hamiltonian. 
In Section 3 the known solutions~\cite{Smilga, Sidorov1, Sidorov2} for one-dimensional
SU$(2|1)$ mechanics are reproduced. In Section 4 we provide the exact supercharges and 
Hamiltonian for so-called real K\"{a}hler spaces, generalizing the results of \cite{AP1,AP2} to 
SU$(2|1)$ supersymmetry. Section 5 specializes to isotropic spaces and extends the solutions 
found in~\cite{N4mech}. We also present explicit solutions for spheres and pseudospheres. 
A few comments and remarks conclude the paper.

\setcounter{equation}{0}
\section{Supercharges and Hamiltonian}
Our goal is to realize the $su(2|1)$ superalgebra
\bea\label{su21}
&&\big\{ Q^a, \bQ_b  \big\} = \sfrac{\im}{2}\delta_b^a H - \mu I_b^a + \im \mu I_0 \delta_b^a\ ,\quad 
\big\{ Q^a, Q^b  \big\} =\big\{ \bQ_a, \bQ_b  \big\}=0\ , \nn \\
&& \big\{ Q^a ,H \big\} = \big\{ \bQ_a, H  \big\} =0\ , \quad 
\big\{ I_0 ,Q^a  \big\} = \sfrac{\im}{2}Q^a\ , \quad 
\big\{ I_0 ,\bQ_a  \big\} = -\sfrac{\im}{2}\bQ_a\ , \\
&& \big\{ I^{ab},I^{cd} \big\} =-\epsilon^{ac}I^{bd} -\epsilon^{bd}I^{ac}\ , \quad 
\big\{ I^{ab},Q^c \big\} =- \sfrac{1}{2}\big( \epsilon^{ac}Q^b + \epsilon^{bc}Q^a  \big)\ , \quad 
\big\{ I^{ab},\bQ_c \big\} = \sfrac{1}{2}\big( \delta^a_c \bQ^b + \delta^b_c \bQ^a  \big)\ ,\nn
\eea
with a constant deformation parameter~$\mu$
on the $(2n{+}4n)$-dimensional phase space given by $n$ coordinates~$x^i$ and momenta~$p_i$,
with $i=1,\ldots,n$, each of which is accompanied by four fermionic ones 
$\psi^{i a}$ and $\bpsi_{b}^ j=(\psi^{j b})^\dagger$.  
On the cotangent bundle over an $n$-dimensional Riemann manifold, 
the Poisson brackets between the basic variables are defined as
\be\label{PB}
\big\{ x^i, {p}{}_j   \big\}= \delta^{i}_j\ ,\quad
\big\{ {\psi}^{ai}, {\bpsi}_b^j \big\} = \sfrac{\im}{2}\delta^a_b \, g^{ij}\;,\quad
\big\{ {p}_i, {\psi}^{aj} \big\} = \Gamma^j_{ik}{\psi}^{ak}\;,\quad 
\big\{ { p}_i, {\bpsi}_a^{j} \big\} = \Gamma^j_{ik}{\bpsi}_a^{k}\ ,\quad
\big\{ {p}_i, { p}_j \big\} = -2\im R_{ijkm}{\psi}^{a k}{\bpsi}_a^m\ .
\ee
Here, $\Gamma^i_{jk}$ and $R^i_{jkl}$  are the components of the Levi--Civita connection and curvature 
of the metric $g_{ij}(x)$ defined in a standard way as
\be\label{def1}
\Gamma^{k}_{ij} \= \sfrac{1}{2} \; g^{km} 
\left( \partial_i g_{jm}+ \partial_j g_{im} - \partial_m g_{ij} \right)
\und
R^i{}_{jkl} \= \partial_k \Gamma^i_{j l} - \partial_l \Gamma^i_{j k}
+ \Gamma^m_{j l}\; \Gamma^i_{m k}-\Gamma^m_{j k}\; \Gamma^i_{m l}\ .
\ee

For the construction of the supercharges $Q^a$ and $\bQ{}_b$ we make use of 
the full U(2) R-symmetry, combining the two types of prepotentials used in \cite{N4mech}:
\be\label{su21Q}
\begin{aligned}
Q^a &\= p_i \psi^{ia} + \im W_i \psi^{ia} + J^{ac}U_i\psi_c^i 
+ \im F_{ijk} \psi^{ic}\psi_c^j \bpsi^{ka} + \im G_{ijk} \psi^{ia}\psi^{jc}\bpsi_{c}^k\ ,  \\
\bQ_a &\= p_i \bpsi^{i}_a - \im W_i \bpsi^{i}_a - J_{ac}U_i\bpsi^{ic} 
+ \im F_{ijk} \bpsi^{i}_c\bpsi^{jc} \psi^{k}_a + \im G_{ijk} \bpsi^{i}_a\bpsi^{j}_c\psi^{ck}\ .
\end{aligned}
\ee
Here, $\epsilon^{ac}W_i$ and $J^{ac}U_i$ are associated with the U(1) and SU(2) parts of the
R-symmetry, generated by $I_0$ and $I^{ac}$, respectively.
To realize the SU(2) currents~$J^{ac}$, one needs to adjoin additional bosonic spin variables 
$\left\{ u^a, \bu_a|\,a=1,2\right\}$~\cite{FIL} parameterizing an internal two-sphere 
and obeying the brackets
\be\label{PBu}
\left\{ u^a, \bu_b \right\} = - \im \, \delta^a_b\ ,
\ee
in terms of which these currents read
\be\label{su2}
J^{ab} = \sfrac{\im}{2} \left( u^a \bu^b +u^b \bu^a\right) \quad \Rightarrow \quad
\left\{ J^{ab} ,J^{cd} \right\} = -\epsilon^{ac} J^{bd} - \epsilon^{bd} J^{ac} .
\ee
The structure functions  $U_i, W_i, F_{ijk}$ and $G_{ijk}$ entering the supercharges \p{su21Q} 
are, for the time being, arbitrary functions of the $n$ coordinates~$x^i$.
In addition, by construction, $F_{ijk}$ and $G_{ijk}$ are symmetric and anti-symmetric 
over the first two indices, respectively:
\be
F_{ijk} =F_{jik}\ , \qquad G_{ijk} = - G_{jik}\ .
\ee

The requirement that the supercharges \p{su21Q} span the $su(2|1)$ superalgebra \p{su21} 
results in the following equations:
\bea
&& G_{ijk} =0, \qquad F_{ijk}-F_{ikj} =0 \quad \Rightarrow \quad
F_{ijk}=F_{(ijk)} \ ,\label{cond1a} \\
&& \nabla_i F_{jkm}- \nabla_j F_{ikm}\=0\ , \label{cond1b} \\
&&  F_{ikp}g^{pq}F_{jmq}-F_{jkp}g^{pq}F_{imq}+R_{ijkm}\=0 \label{cond1c}
\eea
and
\bea
&& \nabla_i W_j - \nabla_j W_i =0 \quad\textrm{and}\quad
\nabla_i U_j - \nabla_j U_i =0 \qquad \Rightarrow \qquad 
W_i = \partial_i W \quad\textrm{and}\quad U_i = \partial_i U \ ,\label{cond1d} \\
&& \nabla_i U_j -U_i U_j  - F_{ijk} g^{km} U_m \=0\ , \label{cond1e} \\
&& \nabla_i W_j + F_{ijk} g^{km}  W_m +\mu\,g_{ij} \=0\ , \label{cond1f} \\
&&g^{ij} W_i U_j - \mu =0 \qquad \textrm{or}\qquad U_j= 0\ , \label{cond1g}
\eea
where, as usual,
\be
\nabla_i W_j \= \partial_i W_j - \Gamma^k_{ij} W_k \und
\nabla_i F_{jkl} \= \partial_i F_{jkl} -\Gamma^m_{ij} F_{klm}-\Gamma^m_{ik} F_{jlm}-\Gamma^m_{il} F_{jkm}\ .
\ee
Finally, the other generators of the $su(2|1)$ superalgebra acquire the form
\be
H = g^{ij}p_i p_j + g^{ij}\partial_i\!W \partial_j\!W +\sfrac{1}{2}J^2\!g^{ij}\partial_i U \partial_j U 
+ 4 \bigl( \epsilon^{cd}\nabla_i \partial_j\!W {-}\im J^{cd}\nabla_i \partial_j U \bigr)\psi_c^i \bpsi_d^j
-4 \big( \nabla_m F_{ijk} {+}R_{ijkm}  \big)\psi^{ic}\bpsi_c^j \, \psi^{kd}\bpsi_d^m, \label{Ham1}
\ee
\be
 I^{ab} \= J^{ab} + \im g_{ij} \big(  \psi^{ia}\bpsi^{jb} + \psi^{ib} \bpsi^{ja}\big) \und
I_0 \= g_{ij}\psi^{ci}\bpsi^j_c\ , \label{Curr1}
\ee
where the Casimir $J^2=J^{cd}J_{cd}$ plays the role of a coupling constant.
The equation \p{cond1b} qualifies $F_{ijk}$ as a so-called third-rank Codazzi tensor \cite{CodazziT}, 
while \p{cond1c} is the curved WDVV equations~\cite{cWDVV}, and \p{cond1d}--\p{cond1g} are 
the deformed analogs of the curved equations considered in \cite{N4mech} and of the flat 
potential equations discussed in \cite{GLP} and~\cite{KL1}.

Two limiting cases are noteworthy. 
First, putting $W=0$ implies via \p{cond1f} that $\mu=0$, bringing us back to the standard
$\mN{=}\,4, d{=}1$ super Poincar\'{e} algebra -- the case  considered in detail in~\cite{N4mech}. 
The converse is not true: $\mu=0$ admits the simultaneous presence of both $U$ and $W$,
as long as their gradients are orthogonal to each other. 
Second, putting $U=0$ solves \p{cond1e} and \p{cond1g}, and it removes the spin variables
together with their currents $J^{ab}$ from the supercharges, the Hamiltonian and the R-currents.
    
Summarizing, to construct SU$(2|1)$ supersymmetric $n$-dimensional mechanics 
on a Riemannian manifold with metric $g_{ij}$, one has to
\begin{itemize}
\item 
solve the curved WDVV equations \p{cond1b}, \p{cond1c} for the fully symmetric function $F_{ijk}$,
\item 
find the admissible prepotentials $W$ and $U$ as solutions to the equations \p{cond1d}--\p{cond1g}.
\end{itemize}
In the following we shall use this procedure.
To begin with, let us demonstrate how the known particular cases of one-dimensional 
SU$(2|1)$ mechanics fit into our scheme.
Then we shall investigate two special geometries allowing for explicit solutions 
of the curved WDVV equations.

\setcounter{equation}{0}
\section{One-dimensional SU(2$|$1) mechanics}
In the distinguished case of a one-dimensional space the metric is always flat and 
can be fixed to $g_{11}=1$ without loss of generality. Therefore, the curved WDVV equations 
become trivial and put no restrictions on the single remaining component $F_{111}$.

The $n=1$ variant of \p{cond1e}--\p{cond1g} reads
\be
U^{\prime\prime}  - F_{111} U^\prime - {U^\prime}^2 \=0\ ,\qquad
W^{\prime\prime} + F_{111} W^\prime + \mu \=0\ ,\qquad
W^\prime U^\prime - \mu =0 \quad\textrm{or}\quad U^\prime \=0\ ,
\ee
where $^\prime$ means differentiation with respect to the single variable $x^1=x$.
These three equations are not independent.
For $U'\neq0$, the two second-order equations follow from each other via $W'U'=\mu$.
In this generic situation, we have the freedom to freely dial one function. 
The choice of any one structure function determines the other two:
\be\label{sol1a}
F_{111} \= -\frac{W^{\prime\prime}+\mu}{W^\prime} 
\= \frac{U^{\prime\prime}-{U^\prime}^2}{U^\prime}  \und
U' = \mu/W' \quad\textrm{or}\quad W' = \mu/U'\ ,
\ee
\be\label{sol1b}
W' \= -\mu\,\ep^{-F_{11}} \smallint^x \ep^{F_{11}} \und
U' \= -\ep^{F_{11}} / \smallint^x \ep^{F_{11}} 
\qquad\textrm{with}\quad F'_{11}=F_{111}\ .
\ee
The Hamiltonian reads
\be
H \= p^2 + \bigl(W'\bigr)^2  + \sfrac12 J^2 \bigl(U'\bigr)^2 
+ 4 \bigl( \epsilon^{cd} W'' -\im J^{cd} U'' \bigr) \psi_c\bpsi_d 
- 4 F'_{111} \psi^{c}\bpsi_c \, \psi^{d}\bpsi_d\ ,
\ee 
which may be expressed purely in terms of either $W'$, $U'$, or $F_{11}$ via \p{sol1a} or \p{sol1b}.

Three different limits can be taken.
First, $W'=0$ yields $\mu=0$. However, $\mu=0$ admits two disjoint solutions,
\be
W'=0 \quad\textrm{and}\quad U' = -\ep^{F_{11}} / \smallint^x \ep^{F_{11}} \qquad\textrm{or}\qquad
U'=0 \quad\textrm{and}\quad W' \sim \ep^{-F_{11}}\ .
\ee
Second, $U'=0$ removes the spin variables, and the Hamiltonian reduces to
\be
H \= p^2 + \bigl( W'\bigr)^2  + 4\,W'' \psi^{a}\bpsi_a 
+ 4 \bigl( \sfrac{W^{\prime\prime}+\mu}{W^\prime}\bigr)'
\psi^{a}\bpsi_a \, \psi^{b}\bpsi_b\ ,
\ee 
which has been constructed in~\cite{Smilga,Sidorov1}.
Third, $F_{111}=0$ leads to 
\be
W' = -\mu\,(x{-}x_0) \und U' = -1/(x{-}x_0)\ ,
\ee
which has been found in~\cite{Sidorov2}. 
In this case the supercharges become linear in the fermions.

\setcounter{equation}{0}
\section{Real K\"{a}hler spaces}
Once we start to consider the $n$-dimensional mechanics, the first problem is to solve the curved WDVV equations
\p{cond1a}--\p{cond1c}. The general solution of these equations is unknown, but in one exceptional case
the solution can easily be constructed. This concerns the so-called `real K\"{a}hler spaces' \cite{AP1,AP2}, 
which are defined by a metric of the form
\be\label{rkm}
g_{ij} \= \frac{\partial^2 G}{\partial x^i \partial x^j} \qquad\Rightarrow\qquad 
\Gamma_{ijk} \= \sfrac{1}{2} \frac{\partial^3 G}{\partial x^i \partial x^j \partial x^k}
\ee
determined by a scalar function~$G$.
It is rather easy to check that two solutions of the curved WDVV equations for such a metric are
\be\label{rkmwdvv}
F^{(1)}_{ijk} \= \Gamma_{ijk} \und F^{(2)}_{ijk} \= -\Gamma_{ijk}\ .
\ee
With this input the equations \p{cond1d}--\p{cond1g} drastically simplify and  can be solved explicitly as
\bea
&& W^{(1)}\=-\mu\,G + \lambda_i x^i \und 
U^{(1)}\=-\log \left( \sigma^i \partial_i G\right), \label{rkmsol1} \\
&& W^{(2)}\=-\mu\left( x^i \partial_i G -G\right) + \lambda^i \partial_i G \und
U^{(2)}\=-\log \left( \sigma_i x^i\right), \label{rkmsol2}
\eea
where $\lambda_i$ and $\sigma^j$ are constants subject to the condition
\be
\sigma_i \lambda^i =0\ .
\ee
Thus, we have a family of $n$-dimensional SU$(2|1)$ mechanics defined on any real K\"{a}hler space.

\setcounter{equation}{0}
\section{Isotropic spaces}
In \cite{cWDVV} a large class of solutions to the curved WDVV equations \p{cond1a}--\p{cond1c} has been constructed on isotropic spaces. 
The metric of such a manifold is SO($n$) invariant, 
i.e.\ it admits $\sfrac12n(n{-}1)$ Killing vectors and can be written in the form
\be\label{CFlat1}
g_{ij} \= \frac{1}{f(r)^2}\,\delta_{ij} \quad\textrm{with}\quad r^2 = \delta_{ij} x^i x^j \qquad\Rightarrow\qquad
\Gamma^k_{ij} \=-\frac{f'}{ r f} \left( x_i \delta_j^k+x_j\delta_i^k-x^k \delta_{ij}\right) 
\ee
with a positive real function~$f$,
where (in this section) $\prime$ means differentiation with respect to $r$.
The ansatz 
\be\label{CFanz1}
F_{ijk} \= a(r)\,x^{i}x^{j}x^{k} + b(r)\bigl(\delta_{ij}x^{k}+\delta_{jk}x^{i}+\delta_{ik}x^{j}\bigr) + f(r)^{-2} F^{(0)}_{ijk}
\ee
extending an arbitrary solution $F^{(0)}_{ijk}$ of the flat WDVV equations~\p{introWDVV}
obeys the curved WDVV equations if \ $x^i F^{(0)}_{ijk}=\delta_{jk}$,
\be\label{solWDVV2}
a \= \frac{2 f \bigl( f- r f'\bigr) \pm \left( 2 f^2 - 3 r f f'+ r^2 (f')^2+ r^2 f f''\right)}{r^4 f^3 \bigl( f - r f'\bigr)}
\quad\und\quad
b \= -\frac{f \pm \bigl( f - r f'\bigr)} {r^2 f^3} \ .
\ee

If we choose the minus sign in the above expressions, i.e.\ for
\be\label{sol11}
a \= \frac{ f f' - r (f')^2 -r f f''}{r^3 f^3 \left( f - r f'\right)} \und b \= - \frac{f'}{r f^3}\ ,
\ee
then a prepotential $W$ solving \p{cond1f} is easily constructed,
\be
W \= w(r,\mu) + W^{(0)} \qquad\textrm{with}\qquad
w'(r,\mu) \= \frac{\alpha(f^2)'-\mu\,r}{2\,f\,(f-rf')}\ ,
\ee
where $\alpha$ is some constant and $W^{(0)}$ obeys the flat equation
\be\label{flatpot}
\partial_i \partial_j W^{(0)}  + F^{(0)}_{ijm}\delta^{mn}\partial_n W^{(0)}\=0 
\qquad\textrm{subject to}\quad x^i \partial_i W^{(0)} = \alpha \ .
\ee
This extends the prepotential solution found in~\cite{N4mech} to $\mu\neq0$.
To this configuration one may add a simple solution to \p{cond1e} for a prepotential~$U$ respecting also~\p{cond1g},
\be\label{su21Usol1}
U \= \log \frac{\mu\,f^2}{\mu\,r^2 - 2\alpha f^2}\ .
\ee
The prepotentials $W$ and $U$ above generate in the Hamiltonian the bosonic potential
\be\label{su21WUpot}
V \= f^2\,\partial_i W^{(0)}\partial_i W^{(0)} 
+ \frac{(\mu\,r - 2\alpha f f^\prime )( \mu\,r^2 - 4\alpha f^2 +2\alpha r f f^\prime )}{4\,r\,(f- r f^\prime)^2} 
+ 2 J^2\,\frac{r^2 \mu^2 (f -r f^\prime )^2}{(\mu\,r^2 - 2 \alpha f^2)^2}\ .
\ee

An interesting case is the (pseudo)sphere, $f=1+\epsilon\,r^2$ with $\epsilon = \pm 1$.
For this manifold, the potential reads
\be\label{su21Wpotsph}
V_{\textrm{sphere}}\= 
(1{+}\epsilon r^2)^2\,\partial_i W^{(0)}\partial_i W^{(0)}
+\frac{(\mu{-}8\epsilon\alpha)^2}{16\epsilon}\,V_{\textrm{Higgs}}
-\frac{\mu^2}{16 \epsilon}
+\frac{8 \epsilon \mu^2\,J^2\,( V_{\textrm{Higgs}}{-}1)}
{\bigl( 8\alpha \epsilon\,V_{\textrm{Higgs}}+\mu\,(1{-} V_{\textrm{Higgs}})\bigr)^2}
\ee
with the Higgs-oscillator potential~\cite{higgs} 
\be
V_{\textrm{Higgs}} \= \biggl( \frac{1+\epsilon r^2}{1-\epsilon r^2}\biggr)^2\ .
\ee 
For $J^2=0$ or $\mu = 8\alpha\epsilon$, simplifications occur,
\be
V_{\textrm{sphere}}\big|_{\mu=8 \alpha \epsilon} \= 
(1{+}\epsilon r^2)^2\,\partial_i W^{(0)}\partial_i W^{(0)} - 4 \alpha^2 \epsilon 
+ 8\epsilon\,J^2\, (V_{\textrm{Higgs}}{-}1)\ .
\ee

\setcounter{equation}{0}
\section{Conclusions}
We extended the previous analysis \cite{N4mech} of $\mN{=}\,4$ supersymmetric mechanics 
on arbitrary Riemannian spaces to systems from $\mN{=}\,4,d{=}1$ super Poincar\'e symmetry 
to SU$(2|1)$ supersymmetry. The extension is parametrized by a deformation parameter~$\mu$,
which only enters in the equation determining the prepotential~$W$ and relating it with 
the prepotential~$U$. All other equations, in particular the curved WDVV equations~\cite{cWDVV}, 
and the form of the supercharges, R-currents and Hamiltonian are unchanged.

A novel feature in our consideration is the presence of both types of prepotentials, $W$ and~$U$,
associated with the U(1) and SU(2) parts of the R-symmetry, respectively.\footnote{
This is actually also possible in the super Poincar\'e limit, but requires their gradients to be mutually orthogonal.}

Two special geometries have been considered in detail. 
Real K\"{a}hler spaces admit an explicit solution for all structure functions. 
On isotropic spaces, we constructed admissible structure functions 
for any conformally invariant solution to the flat structure equations. 
As an application, a Hamiltonian potential for SU$(2|1)$ supersymmetric mechanics 
on a (pseudo)sphere was presented.  All known one-dimensional systems enjoying 
SU$(2|1)$ supersymmetry~\cite{Sidorov1,Sidorov2} can be easily reproduced in our framework. 

One future task even on flat space is a classification of admissible potentials when both prepotentials,
$W$ and~$U$, are present.
At the moment we can do this only for the special case when one of them depends on~$r$ only.
Another interesting question is whether there exist other geometries besides the real K\"{a}hler case
which admit a fully explicit solution.
Since the real K\"{a}hler spaces unambiguously arise in the superfield approach~\cite{AP1,AP2},
it seems compelling to perform a superspace description of the mechanics presented here.
To this end, it is unclear whether the standard superspace is sufficient or whether 
we have to employ the deformed one introduced and advocated in~\cite{Sidorov1,Sidorov2}.

\section*{Acknowledgements}
This work was partially supported by the Heisenberg-Landau program. 
The work of N.K. and S.K. was partially supported by RSCF grant 14-11-00598, 
the one of A.S. by RFBR grant 15-02-06670.  This article is based upon work from COST
Action MP1405 QSPACE, supported by COST (European Cooperation in Science and Technology).

\end{document}